\documentclass[preprintnumbers, floatfix, preprintnumbers, letterpaper, superscriptaddress,nofootinbib, twocolumn]{revtex4}
\pdfoutput=1
\usepackage{graphicx}
\usepackage{microtype}
\usepackage{amsmath}
\usepackage{amssymb}
\usepackage{subfigure}
\usepackage{hyperref}
\usepackage{url}
\usepackage{xcolor}
\usepackage{color}
\usepackage{mathrsfs}
\usepackage{calrsfs}
\usepackage{amsfonts}
\usepackage{latexsym}
\usepackage{ragged2e}
\usepackage{epsfig}
\usepackage{textcomp}
\usepackage{phaistos}
\makeatletter
\renewcommand\@makefnmark{\hbox{\@textsuperscript{\normalfont\color{purple}\@thefnmark}}}
\renewcommand\@makefntext[1]{%
  \parindent 1em\noindent
            \hb@xt@1.8em{%
                \hss\@textsuperscript{\normalfont\@thefnmark}}#1}
\makeatother

\usepackage{caption}
\DeclareCaptionJustification{justified}{\leftskip=0pt \rightskip=0pt \parfillskip=0pt plus 1fil}
\definecolor{vividviolet}{rgb}{0.62, 0.0, 1.0}
\definecolor{amaranth}{rgb}{0.9, 0.17, 0.31}
\definecolor{palatinateblue}{rgb}{0.15, 0.23, 0.89}
\definecolor{brightpink}{rgb}{1.0, 0.0, 0.5}
\definecolor{cornflowerblue}{rgb}{0.39, 0.58, 0.93}
\definecolor{deepcarminepink}{rgb}{0.94, 0.19, 0.22}
\definecolor{radicalred}{rgb}{1.0, 0.21, 0.37}

\hypersetup{ linktoc=all,
    colorlinks, linkcolor={palatinateblue},
    citecolor={brightpink}, urlcolor={amaranth}
}

\graphicspath{{Images/}}

\graphicspath{{Images/}}



\def\sideremark#1{\ifvmode\leavevmode\fi\vadjust{\vbox to0pt{\vss
 \hbox to 0pt{\hskip\hsize\hskip1em
 \vbox{\hsize1.5cm\tiny\raggedright\pretolerance10000
 \noindent #1\hfill}\hss}\vbox to8pt{\vfil}\vss}}}%
\begin{document}

\title{Landauer's principle in Qubit-Cavity quantum-field-theory interaction \\in vacuum and thermal states}

\author{Hao Xu}
\email{haoxu@yzu.edu.cn}
\affiliation{Center for Gravitation and Cosmology, College of Physical Science and Technology, Yangzhou University, \\180 Siwangting Road, Yangzhou City, Jiangsu Province  225002, China}

\author{Yen Chin \surname{Ong}}
\email{ycong@yzu.edu.cn}
\affiliation{Center for Gravitation and Cosmology, College of Physical Science and Technology, Yangzhou University, \\180 Siwangting Road, Yangzhou City, Jiangsu Province  225002, China}
\affiliation{Shanghai Frontier Science Center for Gravitational Wave Detection, Shanghai Jiao Tong University, Shanghai 200240, China}

\author{Man-Hong \surname{Yung}}
\email{yung@sustech.edu.cn}
\affiliation{Department of Physics, Southern University of Science and Technology, Shenzhen 518055, China}
\affiliation{Shenzhen Institute for Quantum Science and Engineering, Southern University of Science and Technology, Shenzhen 518055, China}
\affiliation{Guangdong Provincial Key Laboratory of Quantum Science and Engineering, Southern University of Science and Technology, Shenzhen 518055, China}
\affiliation{Shenzhen Key Laboratory of Quantum Science and Engineering, Southern University of Science and Technology, Shenzhen 518055, China}

\begin{abstract}
Landauer's principle has seen a boom of interest in the last few years due to the growing interest in quantum information sciences. However, its relevance and validity in the contexts of quantum field theory (QFT) remain surprisingly unexplored. In the present paper, we consider Landauer's principle in qubit-cavity QFT interaction perturbatively, in which the initial state of the cavity QFT is chosen to be a vacuum or thermal state. In the vacuum case, the QFT always absorbs heat and jumps to excited states. For the qubit at rest, its entropy decreases, whereas if the qubit accelerates, it may also gain energy and it increases its entropy due to the Unruh effect. For the thermal state, the QFT can both absorb and release heat, depending on its temperature and the initial state of the qubit, and the higher-order perturbations can excite or deexcite the initial state to a higher or lower state. Landauer's principle is valid in all the cases we consider. We hope that this paper will pave the way for future explorations of Landauer's principle in QFT and gravity theories.
\end{abstract}

\maketitle

\emph{Introduction}.---Landauer's principle \cite{landauer1961}, which relates the entropy change of a system to the heat dissipated into a reservoir during any logically irreversible computation, provides a theoretical limit of energy consumption throughout the process. According to this limit, an observer needs at least $k_BT_R\ln2$ of work to erase a one-bit memory, where $k_B$ is the Boltzmann constant and $T_R$ is the temperature of the reservoir at which the erasure process takes place. Landauer's principle provides a direct link between information theory and thermodynamics, and as a consequence, establishes that \emph{information is physical} \cite{landauer1996}. However, ever since its conception, Landauer's principle has been controversial, both theoretically and experimentally \cite{Alicki2012,Orlov12012}. For example, debates ensued on whether the second law of thermodynamics is the premise of Landauer's principle or its outcome, and whether it can be used to exorcise the infamous Maxwell's demon \cite{Earman1999,Bennett2003,0707.3400,Miller2020,Norton2005}. See also the review \cite{Ciliberto2018}.

These seemingly contradictory results arose because they are based on specific models and different (perhaps arguable) assumptions. 
A landmark progress was achieved in 2013 by Reeb and Wolf, who proposed a general and minimal setup to tighten Landauer's principle \cite{Reeb2013} using a quantum statistical physics approach.
Their version of the principle is based on four assumptions: (i) both the “system” $S$ and “reservoir” $R$ are described by Hilbert spaces, (ii) $R$ is initially in a thermal state, (iii) $S$ and $R$ are initially uncorrelated, and (iv) the process  proceeds by unitary evolution. If all the four assumptions are satisfied, then Landauer's principle can be expressed as 
\begin{equation}
\Delta Q\geqslant T_R\Delta S.
\label{bound}
\end{equation}
The quantity $\Delta Q :=\text{tr}\left[\hat{H}_R(\rho'_R-\rho_R) \right]$ is the heat transferred to the reservoir $R$, where $\hat{H}_R$ is the Hamiltonian of $R$, while $\rho'_R$ and $\rho_R$ denote the final and initial state of $R$ respectively, and $\Delta S := S(\rho_S)-S(\rho'_S)$ is the von Neumann entropy change between the initial state $\rho_S$ and the finial state $\rho'_S$ of the system $S$.

The derivation of Reeb and Wolf is simple and illuminating. In particular, it makes use of the non-negativity of two basic quantities in quantum information: \emph{relative entropy} and \emph{mutual information}. The bound \eqref{bound} is written only in terms of $\Delta S$ and $\Delta Q$, and it does not require any information beyond the four assumptions; the bound is also valid arbitrarily far from equilibrium. Due to the growing interest in quantum information sciences, the study of Landauer's principle has picked up the pace, especially in improving (tightening or generalizing) the bound, and carrying out experimental demonstrations in the microscopic domain \cite{1408.5089,Goold2015,Lorenzo2015,Yan2018,Lochan2020,Timpanaro2020}.

On the other hand, the study of Landauer's principle in the areas of quantum field theory (QFT) is surprisingly scarce. 
This is probably because Landauer's principle is born out of information science, while QFT is traditionally more concerned about field interactions. 
However, in recent years, quantum information theory has become interdisciplinary. For example, it was argued that
quantum error correction plays important roles in gauge/gravity correspondence \cite{1411.7041,1503.06237} (and even in the context of a certain two-dimensional conformal field theory \cite{2009.01236}). 
Also, in the attempt to resolve the information paradox of black holes \cite{hawking}, various works have looked into the possible information content of Hawking radiation (see \cite{1409.1231} for a review). Interestingly, the non-negativity of relative entropy -- a crucial property in establishing Landauer's principle -- can also be used to establish \cite{casini} the modern version of the Bekenstein bound (essentially how much energy can be contained by a finite area), which is attained by black holes \cite{bek}. Recently, Landauer's principle has been investigated in general relativity and its implications for gravitational radiation discussed \cite{herrera}.

The application of quantum information to gravity calls for a proper understanding of QFT in a curved spacetime background, in which various quantities (such as temperature) become observer dependent.
Of course, even for QFT in its vacuum state of Minkowski spacetime, an observer with a uniform acceleration will find him or herself in a bath of thermal distribution, with a temperature proportional to the acceleration. This is known as the Unruh effect. Meanwhile, the inertial observer detects nothing peculiar \cite{Unruh1976,Unruh1983}. 

Since the derivation of Reeb and Wolf and its modifications have not considered the models in QFT and observer-dependent effects, in the present work we take the first step by considering two models. One is an accelerating qubit interacting with the vacuum state of free massless scalar QFT, while the other is a qubit at rest interacting with the QFT in a thermal state. Since practical information processing requires finite time erasure and any device designed to perform this task also needs to be built on a finite size platform, in both cases we shall restrict our study to the qubits that interact with the QFT in a finite time and in a finite size cavity. All the four assumptions of Reeb and Wolf are satisfied in both models. We will calculate perturbatively the variations of the von Neumann entropy of the qubits and the heat dissipation into the cavity QFT, and then we check whether the bound \eqref{bound} is still valid. Henceforth in the present work we adopt the natural unit system, setting $c=\hbar=k_B=1$ in all the analytical calculations and numerical analyses. \\

\emph{Detector-cavity QFT interaction}.--- The total Hamiltonian $\hat{H}_{\text{total}}$ describing our system consists of three terms: $\hat{H}_{\text{total}}=\hat{H}^{(d)}_0+\hat{H}^{(f)}_0+\hat{H}_{\text{int}}$. The first term $\hat{H}^{(d)}_0$ is the free Hamiltonian of the detector, and in our case it is just a qubit so we can choose $\hat{H}^{(d)}_0=\Omega_{\text{d}} |e\rangle\langle e|$, where $|e\rangle$ denotes the excited state of the qubit and $\Omega_{\text{d}}$ is the energy level. The second term $\hat{H}^{(f)}_0=\sum_{j=1}^{\infty}\omega_j a^{\dag}_ja_j$ is the free Hamiltonian of the cavity QFT, and finally $\hat{H}_{\text{int}}=\lambda \chi(\tau)\mu(\tau) \phi[x(\tau)]$ is the interaction Hamiltonian, in which $\lambda$ is a weak coupling constant so that we can apply perturbative method, and $\tau$ denotes proper time. Here $\chi(\tau)$ is the so-called ``switching function'' that controls the interaction, $\mu(\tau)$ is the monopole moment of the detector and $\phi[x(\tau)]$ is the field operator at the position of the detector in the cavity. This model has also been used to build quantum gates for the processing of quantum information \cite{1209.4948} and to study weak equivalence principle \cite{1310.5097,1807.07628}. If we solve the system in the interaction picture, the monopole moment can be expressed as $\mu(\tau)=\sigma^{+}e^{i\Omega_{\text{d}} \tau}+\sigma^{-}e^{-i\Omega_{\text{d}} \tau}$, and $\phi[x(\tau)]$ reads
\begin{align}
\phi[x(\tau)] = \sum_{j=1}^{\infty}\left( a_je^{-i\omega_j t(\tau)} u_j\left[x(\tau)\right]+a^{\dagger}_je^{i\omega_j t(\tau)}u_j^*\left[x(\tau)\right] \right),
\label{int}
\end{align}
where the expression of $u_j\left[x(\tau)\right]$ depends on the boundary conditions of the cavity. The time evolution operator of the system under the interaction Hamiltonian $\hat{H}_{\text{int}}$ from time $\tau=0$ to $\tau=T$ is\footnote{This $T$ should not be confused with temperature.} given by the Dyson series:
\begin{align}
\hat{U}(T,0)=&\openone\underbrace{-i\int^{T}_{0}d\tau \hat{H}_{\text{int}}(\tau)}_{\hat{U}^{(1)}} \\ \notag
&\underbrace{+(-i)^2\int^{T}_{0}d\tau \int^{\tau}_{0}d\tau' \hat{H}_{\text{int}}(\tau)\hat{H}_{\text{int}}(\tau')}_{\hat{U}^{(2)}}+ ...\\ \notag
&\underbrace{+(-i)^n\int^{T}_{0}d\tau ... \int^{\tau^{(n-1)}}_{0}d\tau^{(n)} \hat{H}_{\text{int}}(\tau) ... \hat{H}_{\text{int}}(\tau^{(n)})}_{\hat{U}^{(n)}},
\label{dyson}
\end{align}
so the density matrix at a time $\tau=T$ will be 
\begin{equation}
\rho_{T}\!=\!\big[\openone+\hat{U}^{(1)}+\hat{U}^{(2)}+\mathcal{O}(\lambda^3)\big]\rho_0\big[\openone+\hat{U}^{(1)}+\hat{U}^{(2)}+\mathcal{O}(\lambda^3)\big]^{\dagger},
\end{equation}
and we can write $\rho_T$ order by order as
\begin{equation}
\rho_{T}=\rho^{(0)}_{T}+\rho^{(1)}_{T}+\rho^{(2)}_{T}+\mathcal{O}(\lambda^3),
\end{equation}
where
\begin{align}
\rho^{(0)}_{T}&=\rho_0, \\
\rho^{(1)}_{T}&=\hat{U}^{(1)}\rho_0+\rho_0 \hat{U}^{(1)\dagger}, \\
\rho^{(2)}_{T}&=\hat{U}^{(1)}\rho_0 \hat{U}^{(1)\dagger}+\hat{U}^{(2)}\rho_0+\rho_0 \hat{U}^{(2)\dagger}.
\label{rho}
\end{align}
We are now ready to study Landauer's principle for various initial states of the QFT.\\

\noindent 1.\emph{Vacuum State.} \\

Firstly we choose the initial state of the cavity QFT to be the vacuum $|0\rangle\langle 0|$, where $|0\rangle$ satisfies $a_j|0\rangle =0$ for all positive integers $j$. The initial state of the detector is chosen to be $(1-p)|g\rangle\langle g|+p|e\rangle\langle e|$, where $|g\rangle$ and $|e\rangle$ correspond to the ground state and excited state, respectively, thus the initial state for the total system is $\rho_0=\left[ (1-p)|g\rangle\langle g|+p|e\rangle\langle e| \right] \otimes |0\rangle\langle 0|$. Inserting the interaction Hamiltonian into the Dyson series we obtain all the formulas of $\hat{U}^{(N)}$. For the $\rho^{(1)}_{T}$ term, the $a_j$ from $\hat{U}^{(1)}$ acting on the $|0\rangle\langle 0|$ would be 0. On the other hand, $a^{\dagger}_{j}|0\rangle\langle 0|=|1_j\rangle\langle 0|$, which is an off-diagonal term. So if we take the trace of the field the result would be 0. Similarly both $\sigma^{+}$ and $\sigma^{-}$ acting on  $(1-p)|g\rangle\langle g|+p|e\rangle\langle e|$ can only yield off-diagonal terms, thus the trace of the detector would also be 0. This means $\rho^{(1)}_{T}=0$ and the detector-cavity QFT interaction has no effect on the $\lambda$ order. In the language of QFT, it is just the one point function $\langle 0|\phi(x)|0\rangle=0$.

Next we consider the $\lambda^2$ order term $\rho^{(2)}_{T}$. For the term $\hat{U}^{(1)}\rho_0 \hat{U}^{(1)\dagger}$, since we have $a^{\dagger}_{j}$ and $a_{j}$ on both sides of $|0\rangle\langle 0|$, the vacuum state can be excited, and $\sigma^{\pm}$ can also produce diagonal terms. For both $\hat{U}^{(2)}\rho_0$ and $\rho_0 \hat{U}^{(2)\dagger}$, the $\hat{U}^{(2)}$ operator acts on $|0\rangle\langle 0|$ from one side, so the vacuum could not jump to excited states. After some lengthy computations we arrive at
\begin{align}
\hat{U}^{(1)}\rho_0 \hat{U}^{(1)\dagger}=&\lambda^2 \sum_{j=1}^{\infty}\big[(1-p)|I_{+,j}|^2 |e\rangle\langle e| \\ \notag
&+p|I_{-,j}|^2 |g\rangle\langle g| \big]|1_j\rangle\langle 1_j|,
\label{lambda2a}
\end{align}
and
\begin{align}
\hat{U}^{(2)}\rho_0=\rho_0 \hat{U}^{(2)\dagger}=&-\frac{\lambda^2}{2}\sum_{j=1}^{\infty}\big[(1-p)|I_{+,j}|^2 |g\rangle\langle g| \\ \notag
&+p|I_{-,j}|^2 |e\rangle\langle e| \big]|0\rangle\langle 0|,
\label{lambda2b}
\end{align}
where
\begin{equation}
I_{\pm,j}:=\int^{T}_0 d\tau~e^{i\left[\pm \Omega_{\text d} \tau+\omega_jt(\tau)\right]} u_j\left[x(\tau)\right].
\label{I}
\end{equation}
Here we have already set $\chi(\tau)=1$ for $0\leqslant \tau \leqslant T$. These formulas give the evolution of the total system at $\lambda^2$ order. They are also unitarity preserving. Similar analysis can also be extended to higher order of $\lambda$, and one finds the detector-cavity QFT interaction can only affect the system at even-order of $\lambda$. Unitarity is preserved order by order. The vacuum state can be excited in the order of $\lambda^{2n}~(n=1,2,3...)$, while the contributions from $\lambda^{2n-1}$ vanish. However, we empahsize that this is \emph{not} a general result; it depends on the initial state of the QFT. If the initial state already contains some off-diagonal terms, the odd-order $\lambda^{2n-1}$ interaction may create some diagonal terms and the contributions are not zero. For example, if the initial state is a coherent state, the $\lambda$ order would play the leading role \cite{1209.4948}. In the present work we are considering weak coupling, so we focus on at most the $\lambda^2$ terms and omit higher order ones.

Up to order $\lambda^2$ we can write the density matrix of the total system as 
\begin{equation}
\rho_{T} =\rho_0+\hat{U}^{(1)}\rho_0 \hat{U}^{(1)\dagger}+\hat{U}^{(2)}\rho_0+\rho_0 \hat{U}^{(2)\dagger}.
\end{equation}
Tracing out the field part we get the reduced density matrix of the detector:
\begin{equation}
\rho^d_{T}= (1-p-\delta p)|g\rangle\langle g|+(p+\delta p)|e\rangle\langle e|,
\end{equation}
where 
\begin{equation}
\delta p= \lambda^2 \sum_{j=1}^{\infty} \left((1-p)|I_{+,j}|^2-p|I_{-,j}|^2\right).
\end{equation}
Tracing out the detector part we get the reduced density matrix of the field:
\begin{equation}
\rho^f_{T}= (1-\delta f)|0\rangle\langle 0|+\delta f|1_j\rangle\langle 1_j|,
\end{equation}
where
\begin{equation}
\delta f= \lambda^2 \sum_{j=1}^{\infty} \left(p|I_{-,j}|^2+(1-p)|I_{+,j}|^2\right).
\end{equation}
Notice that the sum above also includes the state $|1_j\rangle\langle 1_j|$. Using the definition of $\Delta S$ and $\Delta Q$ in \eqref{bound}, we have
\begin{equation}
\Delta S= \ln{\left(\frac{1-p}{p}\right)}\lambda^2\sum_{j=1}^{\infty} \left(p|I_{-,j}|^2-(1-p)|I_{+,j}|^2\right)
\label{S1}
\end{equation}
and 
\begin{equation}
\Delta Q = \lambda^2 \sum_{j=1}^{\infty} \left(p|I_{-,j}|^2+(1-p)|I_{+,j}|^2\right)\omega_j.
\label{Q1}
\end{equation}
From \eqref{Q1} we can see that heat dissipation to the field is always non-negative. This brings no surprise since our initial state of the field is vacuum, so it cannot transfer heat to the detector. On the other hand, the sign of $\Delta S$ depends on the explicit values of $p$ and $|I_{\pm,j}|^2$. The $|I_{\pm,j}|^2$ term in turn depends on the boundary conditions and qubit trajectories. For example, $u_j\left[x(\tau)\right]\sim \sin[k_n x(\tau)]$ for the Dirichlet boundary condition, and $u_j\left[x(\tau)\right]\sim e^{i k_n x(\tau)}$ for the periodic boundary condition, up to some normalization constants. For either $\sin[k_n x(\tau)]$ or $e^{i k_n x(\tau)}$, if the detector is located at the position $x(\tau)=\text{constant}$, then $t=\tau$ and $u_j\left[x(\tau)\right]$ gives a constant value. From \eqref{I}, one finds for $I_{-,j}$, as $\Omega_{\text{d}}=\omega_j$, the integrand will just be the value of $u_j\left[x(\tau)\right]$ at $x(\tau)=\text{constant}$. Thus $I_{-,j}$ is proportional to the time $T$. However, if $\Omega_{\text{d}}\neq\omega_j$, or for the case of $I_{+,j}$, the integration gives $\frac{1-e^{i(\pm \Omega_{\text d}+\omega_j)T}}{\pm \Omega_{\text d}+\omega_j}u_j(x)$. For larger values of $\pm \Omega_{\text d}+\omega_j$, the contributions from \eqref{I} becomes smaller. Whenever $\omega_j$ is outside a small neighbourhood of $\Omega_{\text d}$, the noise created by these terms quickly decay. An analogous phenomenon in classical mechanics was reported in \cite{Smith2008}.

\begin{figure}
\begin{center}
\includegraphics[width=0.43\textwidth]{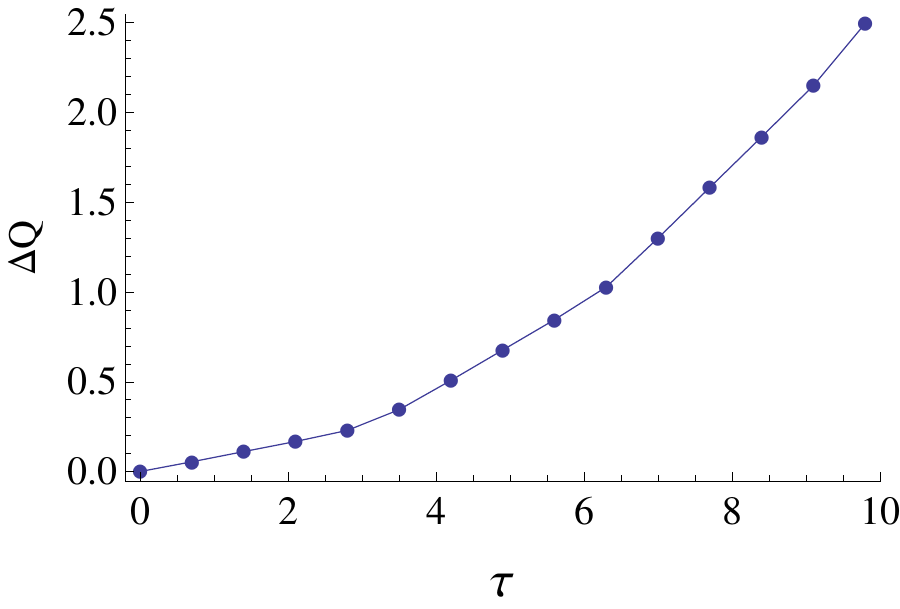}
\includegraphics[width=0.43\textwidth]{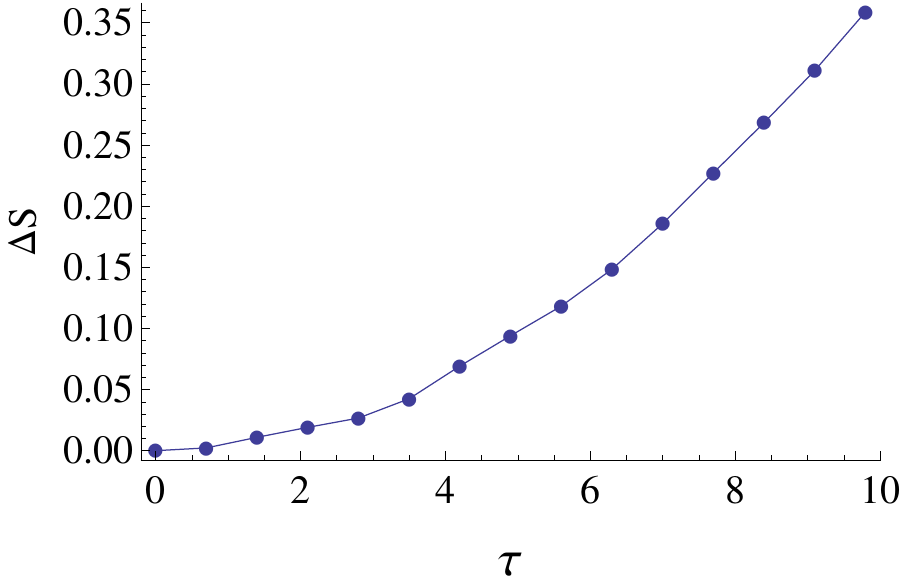}
\caption{Detector at rest for the case of vacuum state. We set $u_j\left[x(\tau)\right]\sim \sin[k_n x(\tau)]$ and the $\Omega_{\text{d}}=\omega_{10}$, $p=0.05$, $L = 1.56789$ and $x=0.212345$ in natural units.}
\label{fig1}
\end{center}
\end{figure}

Thus for the detectors at rest, $I_{+,j}$ is negligible compared to $I_{-,j}$, and both $\Delta S$ and $\Delta Q$ are dominated by the $|I_{-,j}|^2$ term. The vacuum state of the field absorbs heat from the detector, which leads to the decrease in the detector's entropy. Since the effective temperature of the vacuum state is zero, Landauer's principle is satisfied. In FIG.(\ref{fig1}) we present the numerical results of  $\Delta Q$ and $\Delta S$ as the function of $\tau$ for the detector at rest for $p=0.05$. We can observe both $\Delta Q$ and $\Delta S$ increase with the proper time $\tau$. Notice that the settings for the parameters in $|I_{\pm,j}|^2$, such as the cavity scale $L$ and the location of the qubit, need to avoid the possible zeroes of the function due to the periodicity of the integrand. 

\begin{figure}
\begin{center}
\includegraphics[width=0.44\textwidth]{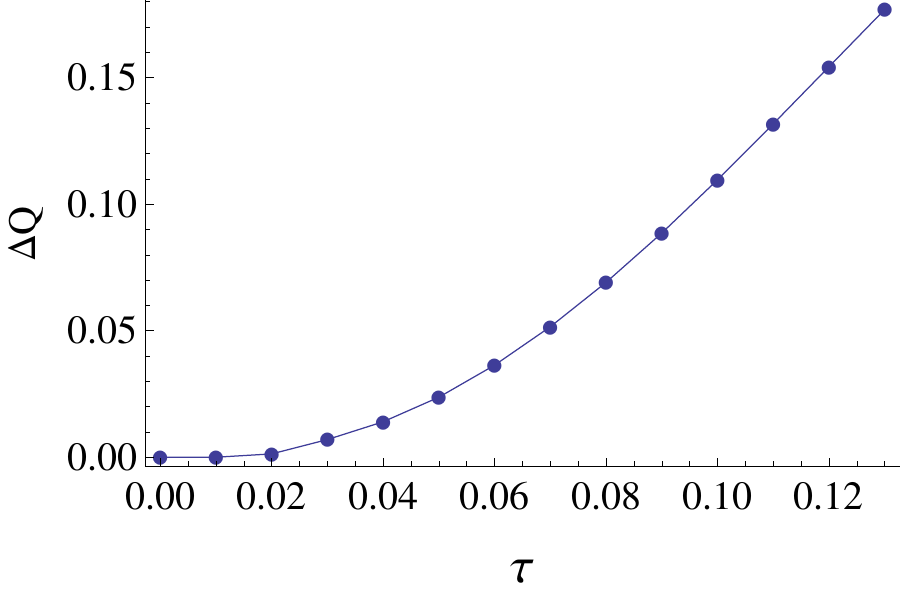}
\includegraphics[width=0.44\textwidth]{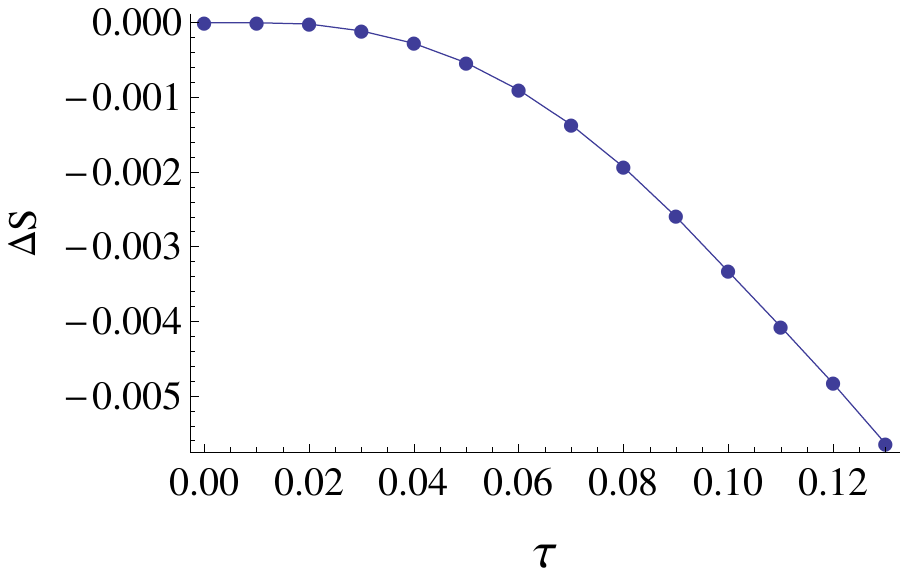}
\caption{Accelerating detector in the case of vacuum state. We set $u_j\left[x(\tau)\right]\sim \sin[k_n x(\tau)]$,  $\Omega_{\text{d}}=\omega_{15}$, $p=0.05$, $a=50$, $L=3$ in natural units.}
\label{fig2}
\end{center}
\end{figure}

For the accelerating detector with a constant proper acceleration $a$, we can choose 
\begin{equation}
x(\tau)=\frac{1}{a}\left(\cosh(a\tau)-1\right), \quad t(\tau)=\frac{1}{a}\sinh(a\tau),
\end{equation}
so that the detector is at $x=0$ at time $t=0$. Inserting the above trajectories into $|I_{\pm,j}|^2$ we find that while $\Delta Q$ is always positive, $\Delta S$ can become negative (see FIG.(\ref{fig2})), meaning both the detector and the field gain energy. The detector seems to be both absorbing photons from the field and emitting photons to the field at the same time. However, since the field is in vacuum, there is no Minkowski photon to be absorbed, and the detector is just emitting photons. This appears to be a violation of energy conservation. There is no real paradox, however: this extra energy comes from the source of the detector acceleration in the first place. The accelerating detector causes the emission of particles that create -- for the lack of a better term -- a ``resistance force'', and the accelerating force has to overcome this resistance by doing more work, which supplies the extra energy \cite{Birrell1982}.

Although the four assumptions of Reeb and Wolf are satisfied in this model, the acceleration of the detector is caused by some other sources, such as an external field or curved spacetime. To obtain the full picture we need more information about the source, and treat it as part of the total system. Nevertheless, we know the bound \eqref{bound} is satisfied, since $\Delta Q>0$ and $T_R=0$, even if $\Delta S$ can in principle be negative. One may fear that $T_R=0$ would render Landauer's principle trivial. However this is because the four assumptions of Reeb and Wolf only provides a general and \emph{minimal} setup. If one wants to improve the bound, extra information of the system is needed, such as an interaction formula \cite{Goold2015} or the heat capacity of the reservoir \cite{Timpanaro2020}. For a specific model one can always include more assumptions to obtain a tighter bound, but this is not the research focus of the present work. In the present work we concentrate on the original bound \eqref{bound} of Reeb and Wolf.  \\

\noindent 2.\emph{Thermal State.} \\

In some sense the vacuum state can be viewed as the vanishing temperature limit of the thermal state, although one of them is pure and the other is mixed. If we consider the initial state of the field to be the thermal state with a non-vanishing temperature $T_R$, it can be written as (this expression is commonly used in the quantum optics community) \cite{Olivares2012}
\begin{equation}
\bigotimes_{j=1}^{\infty}\sum_{n_j=0}^{\infty}\frac{\bar{n}_j^{n_j}}{(1+\bar{n}_j)^{1+n_j}}|n_j\rangle\langle n_j|,
\label{thermal}
\end{equation}
where for each integral value of $j$, $n_j \in [0,\infty)$, and $\bar{n}_j:={1}/{\left(e^{\frac{\omega_j}{T_R}}-1\right)}$. 
Taking the limit $T_R\rightarrow 0$ we have $\bar{n}_j\rightarrow 0$, and the only non-vanishing term would be $n_j=0$, which reduces to the vacuum case. The initial density matrix for the total system is the direct product of $(1-p)|g\rangle\langle g|+p|e\rangle\langle e|$ and \eqref{thermal}. Since \eqref{thermal} contains only diagonal terms, we know the contributions from the odd-order of $\lambda^{2n-1}$ will also be zero for both the detector and the field.

Similarly, we can also calculate $\rho^{(2)}_{T}$. Since now the initial state includes the excited states,  $\hat{U}^{(1)}\rho_0 \hat{U}^{(1)\dagger}$ acting on the $|n_j\rangle\langle n_j|$ state of $\rho_0$ can create both $|n_j+1\rangle\langle n_j+1|$ and $|n_j-1\rangle\langle n_j-1|$.  Upon evaluating $\hat{U}^{(2)}\rho_0$ and $\rho_0 \hat{U}^{(2)\dagger}$, we find they are neither raising nor lowering the state $|n_j\rangle\langle n_j|$. After some lengthy computations we obtain
\begin{align}
&\hat{U}^{(1)}\rho_0 \hat{U}^{(1)\dagger}=\lambda^2\sum_{j=1}^{\infty} \bigg\{\big[(1-p)|I_{+,j}|^2 |e\rangle\langle e| 
\\ \notag
&+p|I_{-,j}|^2 |g\rangle\langle g|\big]\times \sum_{n_j=0}^{\infty}\frac{\bar{n}_j^{n_j}(1+n_j)}{(1+\bar{n}_j)^{n_j+1}}|n_j+1\rangle\langle n_j+1|
\\ \notag
&+\big[(1-p)|I_{-,j}|^2 |e\rangle\langle e|+p|I_{+,j}|^2 |g\rangle\langle g|\big]
\\ \notag
&\times \sum_{n_j=1}^{\infty}\frac{\bar{n}_j^{n_j}n_j}{(\bar{n}_j+1)^{n_j+1}}|n_j-1\rangle\langle n_j-1| \bigg\},
\end{align}
and
\begin{align}
&\hat{U}^{(2)}\rho_0=\rho_0 \hat{U}^{(2)\dagger} =-\frac{\lambda^2}{2}\sum_{j=1}^{\infty}\sum_{n_j=0}^{\infty} \Big[(1-p)\Big(n_j|I_{-,j}|^2
\\ \notag
&+(n_j+1)|I_{+,j}|^2\Big)|g\rangle\langle g|+p\Big(n_j|I_{+,j}|^2
\\ \notag
& +(n_j+1)|I_{-,j}|^2\Big)|e\rangle\langle e|\Big]\times \frac{\bar{n}_j^{n_j}}{(1+\bar{n}_j)^{1+n_j}}|n_j\rangle\langle n_j|.
\end{align}

\begin{figure}
\begin{center}
\includegraphics[width=0.43\textwidth]{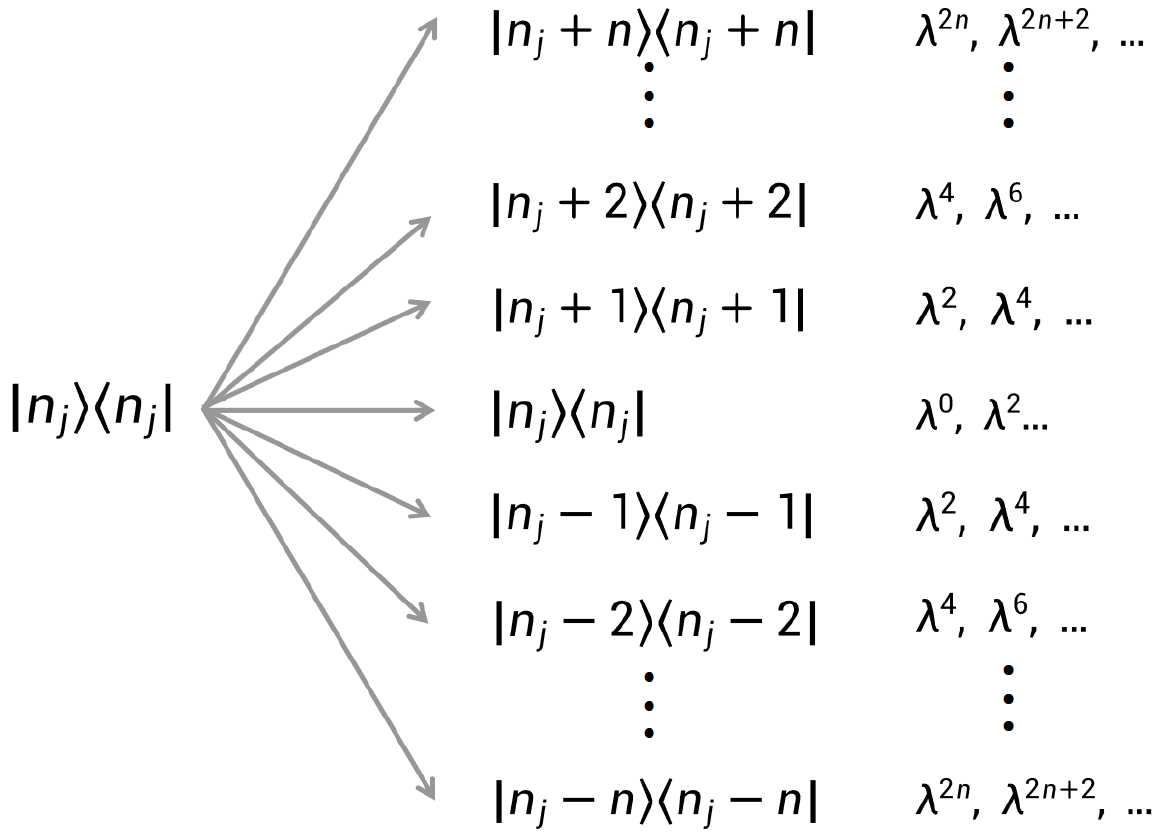}
\caption{The (de-)excition rules for the initial state $|n_j\rangle\langle n_j|$. To create the state $|n_j\pm  n\rangle\langle n_j\pm n|$ one must include the $\lambda^{2n}$ or higher order terms.}
\label{fig3}
\end{center}
\end{figure}

Similar analysis can be extended to the higher order cases. If we consider the $\lambda^{2n}$ order terms, they can be expressed as $\hat{U}^{(m)}\rho_0 \hat{U}^{(2n-m)\dagger}$. For $m=n$ and $m=0, 2n$, we would have $|n_j\pm n\rangle\langle n_j\pm n|$ and $|n_j\rangle\langle n_j|$, respectively, which are exactly what we have obtained in the $n=1$ case above. However, for $n\geqslant 2$ we know $m$ can also take the values from $1$ to $2n-1$. The $\hat{U}^{(m)}\rho_0 \hat{U}^{(2n-m)\dagger}$ term creates $|n_j\pm m\rangle\langle n_j\pm m|$ for $0<m<n$, and $|n_j\pm 2(m-n)\rangle\langle n_j\pm 2(m-n)|$ for $n<m<2n$. Thus, we can conclude that \emph{the $\lambda^{2n}$ order perturbation can create all the states from $|n_j- n\rangle\langle n_j-n|$ to $|n_j+ n\rangle\langle n_j+n|$, as long as $n_j>n$. To create the state $|n_j\pm  n\rangle\langle n_j\pm n|$ one must include the $\lambda^{2n}$ or higher order terms.} In FIG.(\ref{fig3}) we present the (de-)excition rules for the initial state $|n_j\rangle\langle n_j|$.

\begin{figure}
\begin{center}
\includegraphics[width=0.43\textwidth]{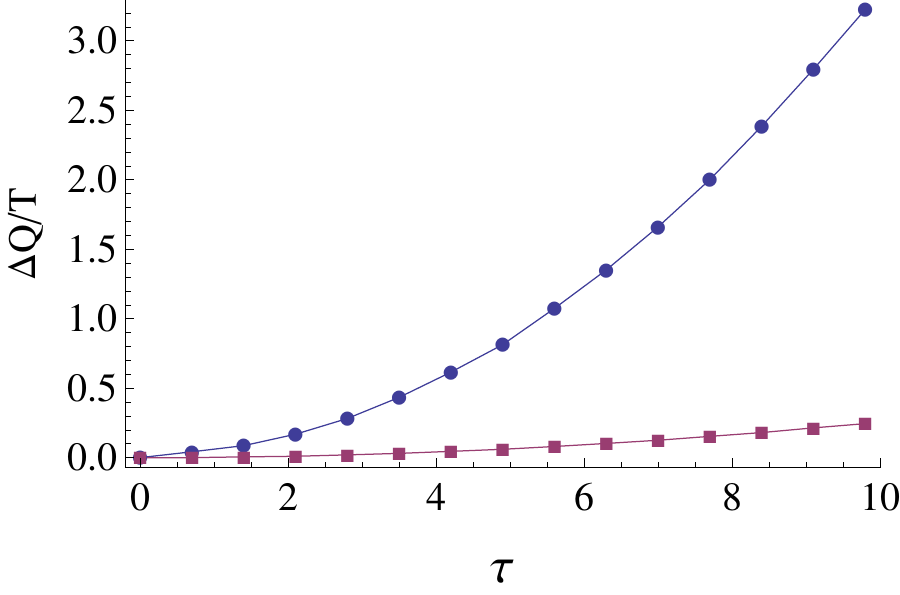}
\includegraphics[width=0.43\textwidth]{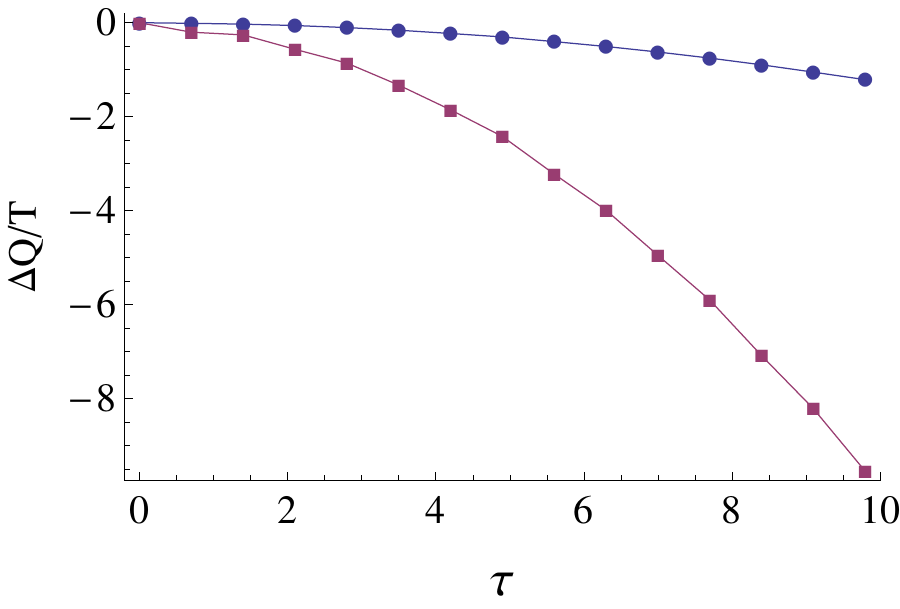}
\caption{The cases of $T_R=1$ and $T_R=100$. In both figures we set $L=1.234$, $\Omega_{\text{d}}=\omega_{15}$, $p=0.05$, $x=0.52345$ in natural units. In each figure the top and bottom curves correspond to ${\Delta Q}/{T_R}$ and $\Delta S$, respectively.}
\label{fig4}
\end{center}
\end{figure}

In the $\lambda^2$ order one can directly check that the total system preserves unitarity. Tracing out the field part and detector part we can obtain the reduced density matrix of the detector and the field, and finally we obtain $\Delta S$ and ${\Delta Q}/{T_R}$ as 
\begin{align}
\Delta S = &\lambda^2\sum_{j=1}^{\infty}\ln{\frac{1-p}{p}}\Big\{[(\bar{n}_j+1)p-\bar{n}_j(1-p)]|I_{-,j}|^2  
\\ \notag
&-[(\bar{n}_j+1)(1-p)-\bar{n}_jp]|I_{+,j}|^2\Big\}, 
\end{align}
and
\begin{align}
\frac{\Delta Q}{T_R}= &\lambda^2\sum_{j=1}^{\infty}\ln{\frac{\bar{n}_j+1}{\bar{n}_j}}\Big\{[(\bar{n}_j+1)p-\bar{n}_j(1-p)]|I_{-,j}|^2  
\\ \notag
&+[(\bar{n}_j+1)(1-p)-\bar{n}_jp]|I_{+,j}|^2\Big\}.
\end{align}
According to Boltzmann distribution, the detector in $(1-p)|g\rangle\langle g|+p|e\rangle\langle e|$ corresponds to an effective temperature $T_d$ satisfying $p=1/(e^{\frac{\Omega_{\text{d}}}{T_d}}+1)$. From the above formulas we can deduce that both $\Delta S$ and $\Delta Q$ are positive for $\frac{\bar{n}_j+1}{\bar{n}_j}>\frac{1-p}{p}$, which means $\frac{T_R}{\omega_j}<\frac{T_d}{\Omega_{\text{d}}}$. This is stronger than the classical condition $T_R<T_d$. Nevertheless, we already know the expressions above are dominated by the $|I_{-,j}|^2$ term in the $\omega_j=\Omega_{\text{d}}$ case, so we also effectively have $T_R<T_d$. In this case we can easily check that Landauer's principle \eqref{bound} is satisfied. Similarly, for $\frac{\bar{n}_j+1}{\bar{n}_j}<\frac{1-p}{p}$, $\Delta S$ and $\Delta Q$ are both negative and \eqref{bound} is also valid. In FIG.\ref{fig4} we present the numerical examples for the $T=1$ and $T=100$ cases.  

\emph{Conclusions}.---In the present work we consider Landauer's principle in qubit-cavity QFT interaction. The initial state of the cavity QFT is chosen to be vacuum or thermal state. In the vacuum case, as the qubit is at rest, the QFT absorbs heat from the qubit and jumps to excited state, while the qubit decreases its entropy. As the qubit accelerates, the qubit may also gain energy and increases its entropy, while the QFT absorbs heat. This extra energy comes from the source of the detector acceleration. In the thermal case, the QFT can both absorb and release heat, depending on its temperature and the qubit's initial state, and the $\lambda^{2n}$ order perturbation can create all the states from $|n_j- n\rangle\langle n_j-n|$ to $|n_j+ n\rangle\langle n_j+n|$, as long as $n_j>n$. In all the cases we consider, Landauer's principle is still valid. Our work thus provides strong support for the effectiveness of Landauer's principle in QFT.

Our analysis can also be extended to other cases in QFT, such as the magnetic field via magnetic dipole moment, which could be more practical in experiment. On the other hand, in the present work we consider weak coupling so that the higher order terms can be neglected. This is because Hamiltonian of the qubit is very different from that of the QFT. However, if we consider a harmonic oscillator to be the detector, the interaction Hamiltonian can be described by the creation-annihilation operators and we can solve the system as a Gaussian state non-perturbatively. This method has been explored in \cite{Brown2013,Bruschi2013} and it would be interesting to study Landauer's principle in harmonic oscillator-cavity QFT interaction. Furthermore, as previously stated, in the case with acceleration the extra energy comes from the acceleration source, so it would be more illuminating to include it as a part of the whole system for a more detailed analysis. If the source comes from gravity, it might even be a first step in shedding some light on the theories of quantum gravity. In fact, the generalized second law is expected to hold in quantum gravity, which in turn implies a quantum singularity theorem \cite{1010.5513}. A better understanding of Landauer's principle might thus eventually lead to a better understanding of the formation of spacetime singularities and the cosmic censorship conjecture.
In addition, recently there are some discussions about the connection between quantum gravity and the gaussianity of state \cite{Howl2021}, we will pursue this direction in our future work.

\begin{acknowledgments}
Hao Xu thanks Karen V. Hovhannisyan, Eduardo Mart\'\i{}n-Mart\'\i{}nez, Yuan Sun and Rui-Hong Yue for useful discussions. He also thanks the Natural Science Foundation of the Jiangsu Higher Education Institutions of China (No.20KJD140001) for funding support. Yen Chin Ong thanks the National Natural Science Foundation of China (No.11922508) for funding support. Man-Hong Yung thanks Natural Science Foundation of Guangdong Province (No.2017B030308003), the Key-Area R\&D Program of Guangdong province (No.2018B030326001), the Science, Technology and Innovation Commission of Shenzhen Municipality (No.JCYJ20170412152620376, No.JCYJ20170817105046702, No.KYTDPT20181011104202253), National Natural Science Foundation of China (No.11875160, No.U1801661), the Economy, Trade and Information Commission of Shenzhen Municipality (No.201901161512), Guangdong Provincial Key Laboratory (No.2019B121203002) for funding support. 
\end{acknowledgments}


\begin{thebibliography}{99}

\bibitem{landauer1961}
R. Landauer, ``Irreversibility and Heat Generation in the Computing Process'', {\hypersetup{urlcolor=vividviolet}\href{https://ieeexplore.ieee.org/document/5392446}{IBM J. Res. Dev. \textbf{5} (1961) 183}}.

\bibitem{landauer1996}
R. Landauer, ``The Physical Nature of Information'', {\hypersetup{urlcolor=vividviolet}\href{https://www.sciencedirect.com/science/article/abs/pii/0375960196004537}{Phys. Lett. A \textbf{217} (1996) 188}}.

\bibitem{Alicki2012}
R. Alicki, ``Quantum Memory as a Perpetuum Mobile? Stability vs. Reversibility of Information Processing'', {\hypersetup{urlcolor=vividviolet}\href{https://www.worldscientific.com/doi/abs/10.1142/S1230161212500163}{Open Systems Information Dynamics \textbf{19} (2012) 1250016}}.

\bibitem{Orlov12012}
A. O. Orlov, C. S. Lent, C. C. Thorpe, G. P. Boechler, G. L. Snider, ``Experimental Test of Landauer's Principle at the Sub-$k_BT$ Level'', {\hypersetup{urlcolor=vividviolet}\href{https://iopscience.iop.org/article/10.1143/JJAP.51.06FE10}{Jpn. J. Appl. Phys. \textbf{51} (2012) 06FE10}}.

\bibitem{Earman1999}
J. Earman, J. D. Norton, ``EXORCIST XIV: The Wrath of Maxwell’s Demon. Part II. From Szilard to Landauer and Beyond'', {\hypersetup{urlcolor=vividviolet}\href{https://www.sciencedirect.com/science/article/abs/pii/S1355219898000264}{Studies in History and Philosophy of Modern Physics \textbf{30} (1999) 1}}.

\bibitem{0707.3400}
H. J. D. Miller, G. Guarnieri, M. T. Mitchison, J. Goold, ``Quantum Fluctuations Hinder Finite-Time Information Erasure Near the Landauer Limit'', {\hypersetup{urlcolor=vividviolet}\href{https://journals.aps.org/prl/abstract/10.1103/PhysRevLett.125.160602}{Phys. Rev. Lett. \textbf{125} (2020) 160602}}.

\bibitem{Norton2005}
J. D. Norton, ``Eaters of the Lotus: Landauer's Principle and the Return of Maxwell’s Demon'', {\hypersetup{urlcolor=vividviolet}\href{https://www.sciencedirect.com/science/article/abs/pii/S1355219804000851}{Studies in History and Philosophy of Modern Physics \textbf{36} (2005) 375}}.



\bibitem{Miller2020}
K. Maruyama, F. Nori, V. Vedral, ``Colloquium: The Physics of Maxwell’s Demon and Information'', 	{\hypersetup{urlcolor=vividviolet}\href{https://journals.aps.org/rmp/abstract/10.1103/RevModPhys.81.1}{Rev. Mod. Phys. \textbf{81} (2009) 1}}.

\bibitem{Bennett2003}
C. H. Bennett, ``Notes on Landauer's Principle, Reversible Computation, and Maxwell's Demon'', {\hypersetup{urlcolor=vividviolet}\href{https://www.sciencedirect.com/science/article/abs/pii/S135521980300039X}{Studies in History and Philosophy of Modern Physics \textbf{34} (2003) 501}}.

\bibitem{Ciliberto2018}
S. Ciliberto, E. Lutz, ``The Physics of Information: From Maxwell to Landauer'', In: Lent C., Orlov A., Porod W., Snider G. (eds) \emph{Energy Limits in Computation}. Springer, Cham, 2018.

\bibitem{Reeb2013}
D. Reeb, M. M. Wolf, ``An Improved Landauer Principle With Finite-Size Corrections'', {\hypersetup{urlcolor=vividviolet}\href{https://iopscience.iop.org/article/10.1088/1367-2630/16/10/103011}{New J. Phys. \textbf{16} (2014) 103011}}.

\bibitem{1408.5089}
Y. Jun, M. Gavrilov, J. Bechhoefer, ``High-Precision Test of Landauer's Principle in a Feedback Trap'', {\hypersetup{urlcolor=vividviolet}\href{https://journals.aps.org/prl/abstract/10.1103/PhysRevLett.113.190601}{Phys. Rev. Lett. \textbf{113} (2014) 190601}}.

\bibitem{Goold2015}
J. Goold, M. Paternostro, K. Modi, ``Nonequilibrium Quantum Landauer Principle'', {\hypersetup{urlcolor=vividviolet}\href{https://journals.aps.org/prl/abstract/10.1103/PhysRevLett.114.060602}{Phys. Rev. Lett. \textbf{114} (2015) 060602}}.

\bibitem{Lorenzo2015}
S. Lorenzo, R. McCloskey, F. Ciccarello, M. Paternostro, G. M. Palma, ``Landauer's Principle in Multipartite Open Quantum System Dynamics'', {\hypersetup{urlcolor=vividviolet}\href{https://journals.aps.org/prl/abstract/10.1103/PhysRevLett.115.120403}{Phys. Rev. Lett. \textbf{115} (2015) 120403}}.

\bibitem{Yan2018}
L. L. Yan, T. P. Xiong, K. Rehan, F. Zhou, D. F. Liang, L. Chen, J. Q. Zhang, W. L. Yang, Z. H. Ma, M. Feng, ``Single-Atom Demonstration of the Quantum Landauer Principle'', {\hypersetup{urlcolor=vividviolet}\href{https://journals.aps.org/prl/abstract/10.1103/PhysRevLett.120.210601}{Phys. Rev. Lett. \textbf{120} (2018) 210601}}.

\bibitem{Lochan2020}
K. Lochan, H. Ulbricht, A. Vinante, S. K. Goyal, ``Detecting Acceleration-Enhanced Vacuum Fluctuations With Atoms Inside a Cavity'', {\hypersetup{urlcolor=vividviolet}\href{https://journals.aps.org/prl/abstract/10.1103/PhysRevLett.125.241301}{Phys. Rev. Lett. \textbf{125} (2020) 241301}}.

\bibitem{Timpanaro2020}
A. M. Timpanaro, J. P. Santos, G. T. Landi, ``Landauer's Principle at Zero Temperature'', {\hypersetup{urlcolor=vividviolet}\href{https://journals.aps.org/prl/abstract/10.1103/PhysRevLett.124.240601}{Phys. Rev. Lett. \textbf{124} (2020) 240601}}.

\bibitem{1411.7041}
A. Almheiri, X. Dong, D. Harlow, ``Bulk Locality and Quantum Error Correction in AdS/CFT'', {\hypersetup{urlcolor=vividviolet}\href{https://link.springer.com/article/10.1007\%2FJHEP04\%282015\%29163}{JHEP \textbf{04} (2015) 163}}.

\bibitem{1503.06237}
F. Pastawski, B. Yoshida, D. Harlow, J. Preskill, ``Holographic Quantum Error-Correcting Codes: Toy Models for the Bulk/Boundary Correspondence'', 	{\hypersetup{urlcolor=vividviolet}\href{https://link.springer.com/article/10.1007\%2FJHEP06\%282015\%29149}{JHEP \textbf{06} (2015) 149}}.

\bibitem{2009.01236}
A. Dymarsky, A. Shapere, ``Solutions of Modular Bootstrap Constraints From Quantum Codes'', {\hypersetup{urlcolor=vividviolet}\href{https://journals.aps.org/prl/abstract/10.1103/PhysRevLett.126.161602}{Phys. Rev. Lett. \textbf{126} (2021) 161602}}.

\bibitem{hawking}
S. W. Hawking, ``Breakdown of Predictability in Gravitational Collapse'', {\hypersetup{urlcolor=vividviolet}\href{https://journals.aps.org/prd/abstract/10.1103/PhysRevD.14.2460}{Phys. Rev. D \textbf{14} (1976) 2460}}.

\bibitem{1409.1231}
D. Harlow, ``Jerusalem Lectures on Black Holes and Quantum Information'', {\hypersetup{urlcolor=vividviolet}\href{https://journals.aps.org/rmp/abstract/10.1103/RevModPhys.88.015002}{Rev. Mod. Phys. \textbf{88} (2016) 015002}}.

\bibitem{casini}
H. Casini, ``Relative Entropy and the Bekenstein Bound'', {\hypersetup{urlcolor=vividviolet}\href{https://iopscience.iop.org/article/10.1088/0264-9381/25/20/205021}{Class. Quant. Grav. \textbf{25} (2008) 205021}}.


\bibitem{bek}
J. D. Bekenstein, ``Universal Upper Bound on the Entropy-to-Energy Ratio for Bounded Systems'', {\hypersetup{urlcolor=vividviolet}\href{https://journals.aps.org/prd/abstract/10.1103/PhysRevD.23.287}{Phys. Rev. D \textbf{23} (1981) 287}}.

\bibitem{herrera}
Luis Herrera, ``Landauer Principle and General Relativity'', {\hypersetup{urlcolor=vividviolet}\href{https://www.mdpi.com/1099-4300/22/3/340}{Entropy \textbf{22(3)} (2020) 340}}.


\bibitem{Unruh1976}
W. G. Unruh, ``Notes on Black-Hole Evaporation'', {\hypersetup{urlcolor=vividviolet}\href{https://journals.aps.org/prd/abstract/10.1103/PhysRevD.14.870}{Phys. Rev. D \textbf{14} (1976) 870}}.

\bibitem{Unruh1983}
W. G. Unruh, R. M. Wald, ``What Happens When an Accelerating Observer Detects a Rindler Particle'', {\hypersetup{urlcolor=vividviolet}\href{https://journals.aps.org/prd/abstract/10.1103/PhysRevD.29.1047}{Phys. Rev. D \textbf{29} (1984) 1047}}.

\bibitem{1209.4948}
E. Mart\'\i{}n-Mart\'\i{}nez, D. Aasen and A. Kempf, ``Processing Quantum Information with Relativistic Motion of Atoms'', {\hypersetup{urlcolor=vividviolet}\href{https://journals.aps.org/prl/abstract/10.1103/PhysRevLett.110.160501}{Phys. Rev. Lett. \textbf{110} (2013) 160501}}.

\bibitem{1310.5097}
A. Ahmadzadegan, E. Mart\'\i{}n-Mart\'\i{}nez, Robert B. Mann, ``Cavities in Curved Spacetimes: The Response of Particle Detectors'', {\hypersetup{urlcolor=vividviolet}\href{https://journals.aps.org/prd/abstract/10.1103/PhysRevD.89.024013}{Phys. Rev. D \textbf{89} (2014) 024013}}.

\bibitem{1807.07628}
E. Tjoa, Robert B. Mann, E. Mart\'\i{}n-Mart\'\i{}nez, ``Particle Detectors, Cavities, and the Weak Equivalence Principle'', {\hypersetup{urlcolor=vividviolet}\href{https://journals.aps.org/prd/abstract/10.1103/PhysRevD.98.085004}{Phys. Rev. D \textbf{98} (2018) 085004}}.

\bibitem{Smith2008}
S. T. Smith, R. Onofrio,``Thermalization in Open Classical Systems With Finite Heat Baths'', {\hypersetup{urlcolor=vividviolet}\href{https://link.springer.com/article/10.1140/epjb/e2008-00070-8}{Eur. Phys. J. B \textbf{61} (2008) 271}}.

\bibitem{Birrell1982}
N. D. Birrell, P. C. W. Davies, \emph{Quantum Fields in Curved Space}, {\hypersetup{urlcolor=vividviolet}\href{https://www.cambridge.org/core/books/quantum-fields-in-curved-space/95376B0CAD78EE767FCD6205F8327F4C}{Cambridge Monographs on Mathematical Physics (Cambridge Univ. Press, Cambridge, UK, 1982)}}.

\bibitem{Olivares2012}
S. Olivares, ``Quantum Optics in the Phase Space'', {\hypersetup{urlcolor=vividviolet}\href{https://link.springer.com/article/10.1140/epjst/e2012-01532-4}{Eur. Phys. J. Special Topics \textbf{203} (2012) 3}}.

\bibitem{Brown2013}
E. G. Brown, E. Martin-Martinez, N. C. Menicucci, R. B. Mann, ``Detectors for Probing Relativistic Quantum Physics Beyond Perturbation Theory'', {\hypersetup{urlcolor=vividviolet}\href{https://journals.aps.org/prd/abstract/10.1103/PhysRevD.87.084062}{Phys. Rev. D \textbf{87} (2013) 084062}}.

\bibitem{Bruschi2013}
D. E. Bruschi, A. R. Lee, I. Fuentes, ``Time Evolution Techniques for Detectors in Relativistic Quantum Information'', {\hypersetup{urlcolor=vividviolet}\href{https://iopscience.iop.org/article/10.1088/1751-8113/46/16/165303}{J. Phys. A: Math. Theor. \textbf{46} (2013) 165303}}.

\bibitem{1010.5513}
A. C. Wall, ``The Generalized Second Law implies a Quantum Singularity Theorem'', {\hypersetup{urlcolor=vividviolet}\href{https://iopscience.iop.org/article/10.1088/0264-9381/30/16/165003}{Class. Quantum Grav. \textbf{30} (2013) 165003}}.

\bibitem{Howl2021}
R. Howl, V. Vedral, D. Naik, M. Christodoulou, C. Rovelli, A. Iyer, ``Non-Gaussianity as a Signature of a Quantum Theory of Gravity'', {\hypersetup{urlcolor=vividviolet}\href{https://journals.aps.org/prxquantum/abstract/10.1103/PRXQuantum.2.010325}{PRX Quantum \textbf{2} (2021) 010325}}.




\end{thebibliography}
\end{document}